\documentclass[a4paper,11pt]{article}
\pdfoutput=1 

\usepackage{jheppub} 

\usepackage{graphicx}
\usepackage{dcolumn}
\usepackage{bm}
\usepackage[colorlinks=true,
urlcolor=cambridgeblue,
linkcolor=darkraspberry,
citecolor=cambridgeblue,
linktocpage=true,
pdfproducer=medialab,
pdfa=true,
anchorcolor=blue]{hyperref}%

\usepackage{cancel}
\usepackage{amssymb}
\usepackage{textcomp}
\usepackage{amsmath}
\usepackage{mathrsfs}
\usepackage{subfigure}
\usepackage{feynmp-auto}

\usepackage{empheq}
\usepackage{bm}
\usepackage{framed}
\usepackage{pdfpages}

\usepackage{float}
\usepackage[utf8]{inputenc} 
\usepackage{xcolor}
\definecolor{cambridgeblue}{rgb}{0.64, 0.76, 0.68}
\definecolor{darkraspberry}{rgb}{0.53, 0.15, 0.34}

\title{Reconstructing particle masses in events with displaced vertices \\}

\preprint{}

\author{Giovanna Cottin}
\affiliation{Cavendish Laboratory, University of Cambridge, J.J. Thomson Ave, Cambridge CB3 0HE, UK}
\affiliation{Department of Physics, National Taiwan University, Taipei  10617, Taiwan}

\emailAdd{gcottin@phys.ntu.edu.tw}

\date{\today}

\abstract{We propose a simple way to extract particle masses given a displaced vertex signature in event topologies where 
two long-lived mother particles decay to visible particles and an invisible 
daughter. The mother could be either charged or neutral and the neutral daughter could correspond to a dark matter particle 
in different models. The method allows to extract the parent and daughter masses by using on-shell conditions and energy-momentum conservation, in addition to the displaced decay positions of the parents, which
allows to solve the kinematic equations fully on an event-by-event basis. We show the validity of the method by
means of simulations including detector effects. If displaced events are seen in discovery searches at the Large 
Hadron Collider (LHC), this technique can be applied.}

\begin{document}
\maketitle

\section{Introduction} 

Almost all of our models for new physics beyond the Standard Model (SM) are able to describe massive, long-lived particles (LLPs), with macroscopic decays, that could potentially be reconstructed as displaced vertices inside the LHC detectors~\cite{Aaboud:2017iio,Aad:2015rba,Aad:2015uaa,CMS:2014wda,CMS:2014hka,Aaij:2016xmb,Aaij:2017mic}.  LHC searches able to make use of displaced vertices benefit from considerably lower backgrounds than searches which cannot, making them sensitive to very small signals of new physics.  This sensitivity is becoming increasingly important in the light of ongoing null results from LHC; new physics may be so feebly coupled to the SM that it is invisible to searches not dedicated to LLPs (see Ref.~\cite{Curtin:2017quu} for a recent and comprehensive review).  New physics models where the correct dark matter relic abundance is obtained via the FIMP~\cite{Hall:2009bx} or SuperWIMP mechanisms~\cite{Feng:2003uy} predict displaced signatures at colliders~\cite{Bernal:2017kxu,Co:2015pka,Arcadi:2014tsa,Arcadi:2013aba,Chang:2009sv}. Models of pseudo-Dirac dark matter~\cite{Davoli:2017swj} also predicts displaced vertices. It is therefore natural to ask the question of what else could be measured at colliders, and what could thus be inferred about the nature of dark matter, given a displaced vertex signal. 

In this work we address the above by developing a simple method to reconstruct particle masses in events with displaced vertices. Identifying particle masses in decay chains where there is a displaced particle decaying at a known position has previously been suggested in Ref.~\cite{Park:2011vw}, where the authors recover the unknown kinematic quantities by using the constrains from the presence of displaced tracks. Our work goes beyond this by demonstrating that our reconstruction technique is resilient to unavoidable instrumentation effects: namely uncertaintes in the reconstructed positions of displaced vertices, and in the magnitude of reconstructed lepton and jet momenta, jets and missing transverse momentum.   
For resolutions representative of a typical LHC detector, we estimate the precision with which the masses of the long-lived particle and its daughter might be  determined given a straw signal. We draw events for that signal from a point in the simplified model proposed in Reference~\cite{Buchmueller:2017uqu}, wherein candidate for dark matter is present as a decay product (daughter) of the long-lived particle which generates the displaced vertices. We describe the topology and kinematic equations in Section~\ref{kinematics}. Our simulations, description of the method and results are presented in Section~\ref{results}. We conclude in Section~\ref{close}.

\section{Kinematics of displaced events}
\label{kinematics}

\newcommand{\rhat}{\bm{\hat{r}}}
\newcommand{\rhatp}{\bm{\hat{r}'}}
\newcommand{\pv}{{\bm{p_V}}}
\newcommand{\pvp}{{\bm{p_{V'}}}}
\newcommand{\pchi}{{\bm{p_\chi}}}
\newcommand{\pchip}{{\bm{p_\chi'}}}
\newcommand{\pchione}{{\bm{p_{\chi_1}}}}
\newcommand{\pchitwo}{{\bm{p_{\chi_2}}}}
\newcommand{\pchionep}{{\bm{p_{\chi'_1}}}}
\newcommand{\pchitwop}{{\bm{p_{\chi'_2}}}}

\begin{figure}[ht]
\centering
\includegraphics[width=0.5\textwidth,angle=0]{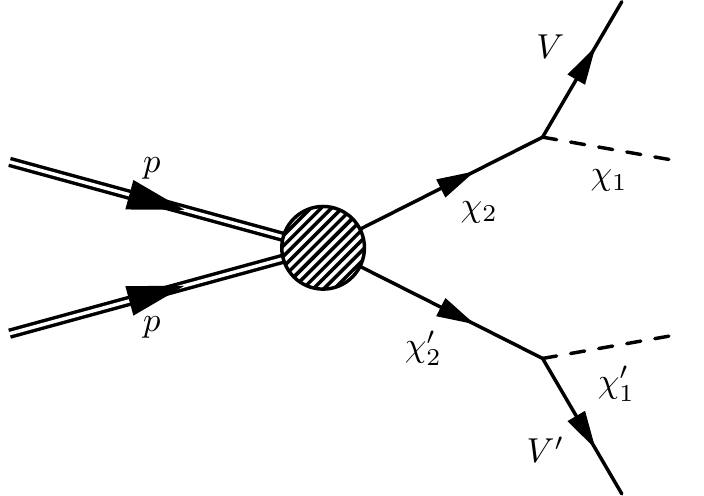}
\caption{Topology of the displaced decays considered in this work. Two long-lived parent particles $\chi_2$ are produced from the primary collision and then decay with a displaced vertex to an invisible daughter $\chi_1$ and something visible $V$, which can be either a single particle or the sum of more particles in the final state.}
\label{MyTopology}
\end{figure}

The event topology considered in this work is shown in Figure~\ref{MyTopology}. It assumes production of a pair of long-lived parent particles, $\chi_2$ and $\chi'_2$, having the same mass as each other.  After moving through displacements $\bm r$ and $\bm r'$, these subsequently decay  to identical invisible daughters $\chi_1$ and $\chi'_1$ in association with visible products $V$ and $V'$ having observed momenta $\bm p_V$ and $\bm p_{V'}$. The accuracy with which the positions $\bm r$ and $\bm r'$ might be experimentally measured will depend on the nature of $V$ and $V'$.  In the most ideal case, $\bm r$ might be reconstructed with a small Gaussian uncertainty as the location from which a two or more charged tracks -- as might happen if $V$ was an $e^+e^-$--pair directly coming from a three-body decay of a neutralino.  If $V$ contained only one visible track, however, it might only be possible to state that  $\bm r$  lives `somewhere on a given track segment'.\footnote{For example, the general direction of a strongly collimated $e^+e^-$ pair coming from a highly boosted $V$ could be well measured, but the actual production point might be only localised to, say, being `somewhere between the third and fourth silicon tracking layer', as hits might have been seen on and beyond the fourth layer but not before.} 

Given the observed event variables $\bm r$, $\bm r'$, $\bm p_V$, $\bm p_{V'}$ and the missing transverse momentum $\bm{p}^{\text{miss}}_{T}$, is it possible to calculate the values of the three-momenta
$\bm p_{\chi_1}$, 
$\bm p_{\chi_2}$, 
$\bm p'_{\chi_1}$ and
$\bm p'_{\chi_2}$ which are compatible with the kinematics of Figure~\ref{MyTopology}?
The answer to this question is ``Yes, if and only if $\bm p_V+ \bm p_{V'}+\bm{p}^{\text{miss}}_{T}$ lies between $\bm r$ and $\bm r'$ in the transverse plane.'' If this condition is not met, the observed event is incompatible with the proposed kinematic structure.  A more mathematical statement and proof of this statement is provided in the Appendix, but the key result noted here is that, for events for which a valid solution exists, the resulting three-momenta
$\bm p_{\chi_1}$, 
$\bm p_{\chi_2}$, 
$\bm p'_{\chi_1}$ and
$\bm p'_{\chi_2}$
are uniquely determined and given by simple algebraic functions of the observed event variables, supplied in the Appendix.

When the three-momenta $\bm p_{\chi_1}$, 
$\bm p_{\chi_2}$, 
$\bm p'_{\chi_1}$ and
$\bm p'_{\chi_2}$ are uniquely determined, the same is not necessarily true for the  masses $m_{\chi_1}$ and $m_{\chi_2}$.
These are constrained only by four-momentum conservation to solve the two equations:
\begin{align}
\left( \sqrt{m_{\chi_2}^2 + \pchitwo^2}, \pchitwo\right)^\mu
-
\left( \sqrt{m_{V}^2 + \pv^2}, \pv \right)^\mu
&=
\left( \sqrt{m_{\chi_1}^2 + \pchione^2}, \pchione\right)^\mu
\qquad{\text{and}}
\\
\left( \sqrt{m_{\chi_2}^2 + \pchitwop^2}, \pchitwop\right)^\mu
-
\left( \sqrt{m_{V'}^2 + \pvp^2}, \pvp \right)^\mu
&=
\left( \sqrt{m_{\chi_1}^2 + \pchionep^2}, \pchionep\right)^\mu,
\end{align}
which upon squaring read\begin{align}
m_{\chi_2}^2+m_V^2 - 2\left( E_V \sqrt{m_{\chi_2}^2 + \pchitwo^2} -  \pchitwo \cdot \pv\right)
&=
m_{\chi_1}^2  \label{eqmuytr}
\qquad{\text{and}}
\\
m_{\chi_2}^2+m_{V'}^2 - 2\left( E_{V'} \sqrt{m_{\chi_2}^2 + \pchitwop^2}  -  \pchitwop \cdot \pvp\right)
&=
m_{\chi_1}^2.
\label{eqfinalSetNew}
\end{align}

In principle, equations~(\ref{eqmuytr}) and (\ref{eqfinalSetNew}) have eight solutions for the mass pair $(m_{\chi_1},m_{\chi_2})$, but we are interested only those resulting in positive masses, of which there are variously zero, one or two solutions per event. 

\section{Simulations and results}
\label{results}

We choose for our study the simplified {\texttt{DisplacedDM}} model that produces displaced vertices plus missing transverse momenta defined in Reference~\cite{Buchmueller:2017uqu}. We consider the model in Figure 3 of~\cite{Buchmueller:2017uqu}, where a quark anti-quark pair decays to a heavy mediator $Y_{1}$ (with spin 1), that then decays to two long-lived parent particles $\chi_{2}$. $\chi_{2}$ decays displaced to a dark matter particle $\chi_{1}$ and a light mediator $Y_{0}$ (with spin 0), that further decays to fermions. The full decay chain of this simplified model is given by

\begin{equation}
q\bar{q}\rightarrow Y_{1} \rightarrow \chi_{2}\bar{\chi}_{2} \rightarrow \chi_{1} Y_{0} \chi_{1} Y_{0} \rightarrow \chi_{1}f\bar{f} \chi_{1}f\bar{f}.
\end{equation}

The authors in Reference~\cite{Buchmueller:2017uqu} provide the corresponding \textsc{UFO}~\cite{Degrande:2011ua} for the {\texttt{DisplacedDM}} model, which we use to simulate events for the process $pp \rightarrow Y_{1} \rightarrow \chi_{2}\bar{\chi}_{2} $ at $\sqrt{s}=13$ TeV using \textsc{MadGraph5\_aMC@NLOv2.5.5}~\cite{Alwall:2014hca} at leading order. The output corresponds to unweighted events in \textsc{LHE} format~\cite{Alwall:2006yp}, that includes the lifetimes of the $\chi_{2}$ particles.\footnote{ The lifetime information can be passed to the \textsc{LHE} events by setting the \texttt{time\_of\_flight} variable in \textsc{MadGraph5}'s \texttt{run\_card}. We choose $1\mathrm{e}^{-25}$ as threshold for displaced vertices.}. 

We set the mass of the mediators to be $m_{Y_{1}} = 1$ TeV and $m_{Y_{0}} = 40$ GeV, so that we can scan combinations for the truth value of the masses $m_{\chi_{1}}$ and $m_{\chi_{2}}$ such that $m_{\chi_2}-m_{\chi_1}\sim\mathcal{O}(10s)$ GeV. We generate a grid of 510 points, with $m_{\chi_{1}}=[1,..,10]$ GeV and $m_{\chi_{2}}=[50,..,100]$ GeV. The size of the lifetime $\tau$ of $\chi_2$ is of the order of $ c\tau \sim 20$ mm, leading to decay lengths of $\mathcal{O}(100)$ mm (after considering the boost factor), which will happen inside the inner trackers of the LHC detectors\footnote{  The ATLAS detector, 
for example, can efficiently reconstruct displaced vertices inside the tracker with decay 
lengths between 4 mm and 300 mm~\cite{Aaboud:2017iio}. In what follows, only displaced decays inside the inner trackers are considered. However, the validity of the method can also be tested if one considers known displaced vertex positions inside the muon spectrometer, as the muon spectrometer also has the capability to reconstruct vertices.}.

The generated events are interfaced to \textsc{Pythia8 v2.15}~\cite{Sjostrand:2014zea} for hadronisation and computation of the $\chi_{2}\rightarrow \chi_{1} f\bar{f}$ decays. The masses and widths of the particles in the model are communicated to \textsc{Pythia} via the SLHA~\cite{Skands:2003cj,Allanach:2008qq} section of the LHE header (which is the same approach adopted in Ref.~\cite{Buchmueller:2017uqu}). We consider the case where $Y_{0}$ decays to $e^{+}e^{-}$ (corresponding to one of the minimal sets of long-lived plus missing transverse momenta benchmarks defined in Table 4 of Ref.~\cite{Buchmueller:2017uqu}).

We first analyze events at the truth-level. We identify the positions of the $\chi_2$ displaced decays in \textsc{Pythia8}, in addition to the 4-momenta of the final state particles from the vertices. We always require the presence of two displaced vertices in each event. This information is further analyzed with {\texttt{python}} routines to solve the kinematic equations numerically. In the case where we have two solutions for the mass pair $(m_{\chi_{1}},m_{\chi_{2}})$ per event, we choose the smallest of the two, which is presumably the correct one (as the mass values can go all the way up, but can never go below zero). Plots are generated with {\texttt{matplotlib}}~\cite{Hunter:2007}.

\begin{figure}[h]
\centering
\includegraphics[width=0.55\textwidth,angle=0]{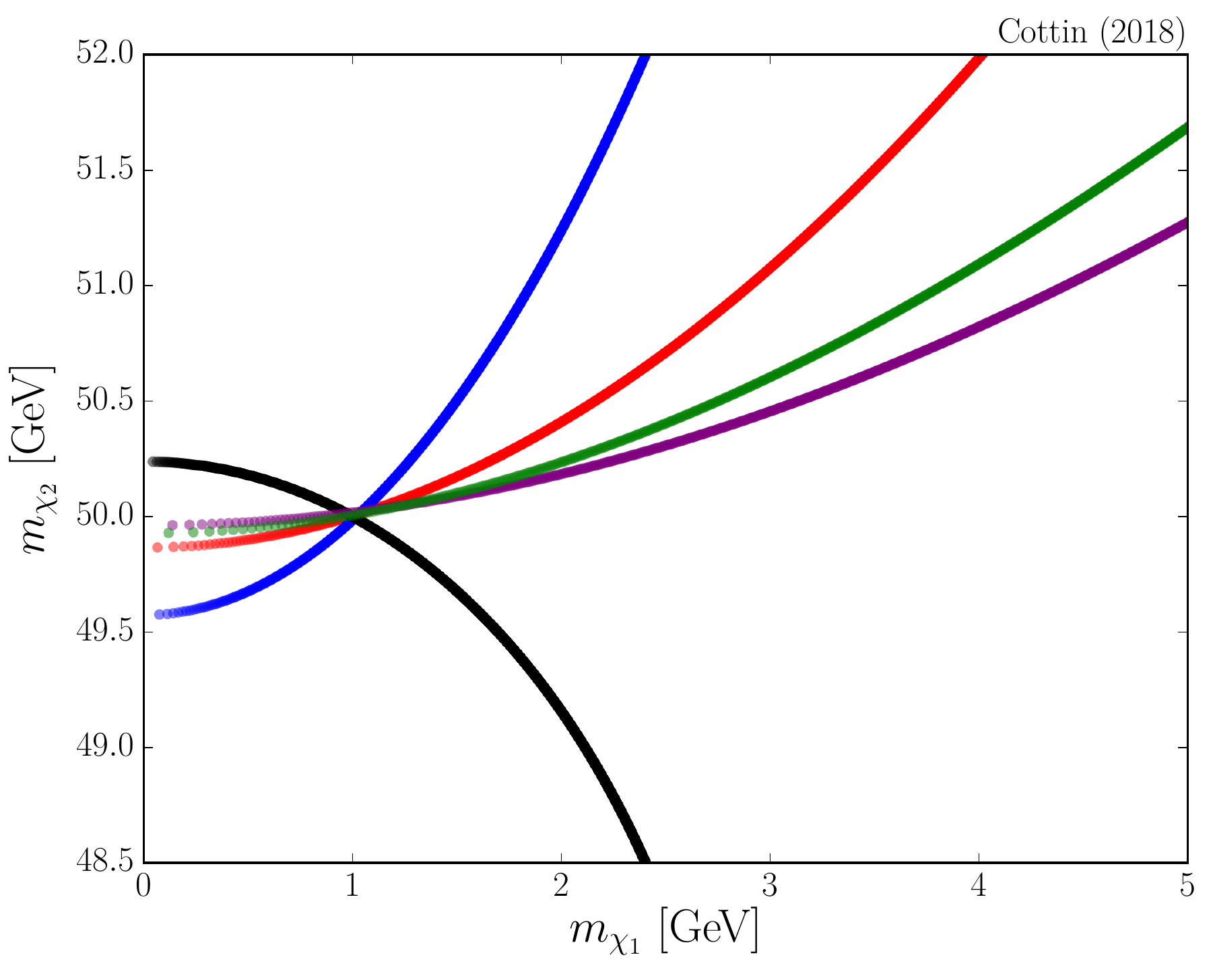}
\caption{Solutions for the masses for 5 events after smearing only one vertex linearly as in equation~(\ref{onevertexSmear}), with $\theta=[-0.1,0.1]$, in the direction of the visible momenta coming out of the vertex. The truth value for the mass pair is $(m_{\chi_1}, m_{\chi_2})= (1, 50)$ GeV.}
\label{figVertexSmear}
\end{figure}

Figure~\ref{figVertexSmear} shows the resulting value of the masses $(m_{\chi_1},m_{\chi_2})$ for 5 events, when one dimensional degree of freedom $\theta$ is added in the direction of the visible momenta coming out of one of the displaced vertices\footnote{Note that $\theta$ should have units of length-over-momentum in principle. We do not address a meaningful size to the spread here, as the purpose is to only show that the kinematic equations and their solutions are correct. A meaningful spread is given once we include all detector effects.}, such that 

\begin{equation}
\bm{r} \longrightarrow \bm{r} + \theta \bm{p_{V}}. 
\label{onevertexSmear}
\end{equation}

By doing this operation, we see the effect on the masses of not knowing about the displaced tracks starting positions. Each curve corresponds to one independent event, and we can see that all of them intersect at the correct truth value for the masses, confirming the equation solving process is accurate.

To take into account inefficiencies in the displaced vertex reconstruction, we smear the positions of the two vertices in each direction by sampling from a Gaussian distribution with $\sigma = 300 \mu$m. Even though displaced vertex reconstruction efficiency gets worse at large radii (specially beyond the pixel layers~\cite{Aad:2015rba}) we use a constant 300 $\mu$m as a conservative choice, as most of our displaced decays occur within transverse position $< 300$ mm. Note that we consider a cylinder with radius $r = 11$ m and length $|z| = 28$ m to model the size of the ATLAS inner detector~\cite{Aad:2009wy}. Any particle that decays outside the inner detector is therefore considered to be stable. 

Leptons, jets and missing transverse momenta are reconstructed inside \textsc{Pythia8}. We use \textsc{FastJet 3.1.3}~\cite{Cacciari:2011ma} for jet reconstruction. The detector response for these objects is modeled in the same way as in Reference~\cite{Allanach:2016pam}, where the jet momentum is smeared by a Gaussian with different resolutions depending on the jet's energy. For electron resolution we use $2\%$ at 10 GeV, falling linearly to $1\%$ at 100 GeV, and then $1\%$ flat. 

For the study of the displaced $e^{+}e^{-}$ system we require at least 4 electrons in each event. Each electron has to be matched to a truth displaced track coming from the displaced vertex. We perform the matching by requiring the distance in the $(\eta, \phi)$ plane between the track and the reconstructed electron to be less than $0.1$. Displaced tracks are defined to have a transverse impact parameter $|d_{0}|>2$ mm and $p_{T}> 1$ GeV, with $d_{0}$ defined in the Appendix of Ref.~\cite{Allanach:2016pam}. For the events that satisfy these requirements we save the smeared momenta for the matched electrons. The missing transverse momenta in each event is also extracted from the detector simulation. The smeared quantities $(\bm{r}, \bm{r'}, \bm{p_{V}}, \bm{p_{V'}}, p^{\text{miss}}_{x}, p^{\text{miss}}_{y})$ are now the new input to our \texttt{python} routine in order to solve the kinematic equations in~(\ref{eqmuytr}) and (\ref{eqfinalSetNew}). 

\begin{figure}[h]
\centering
\includegraphics[width=\textwidth,angle=0]{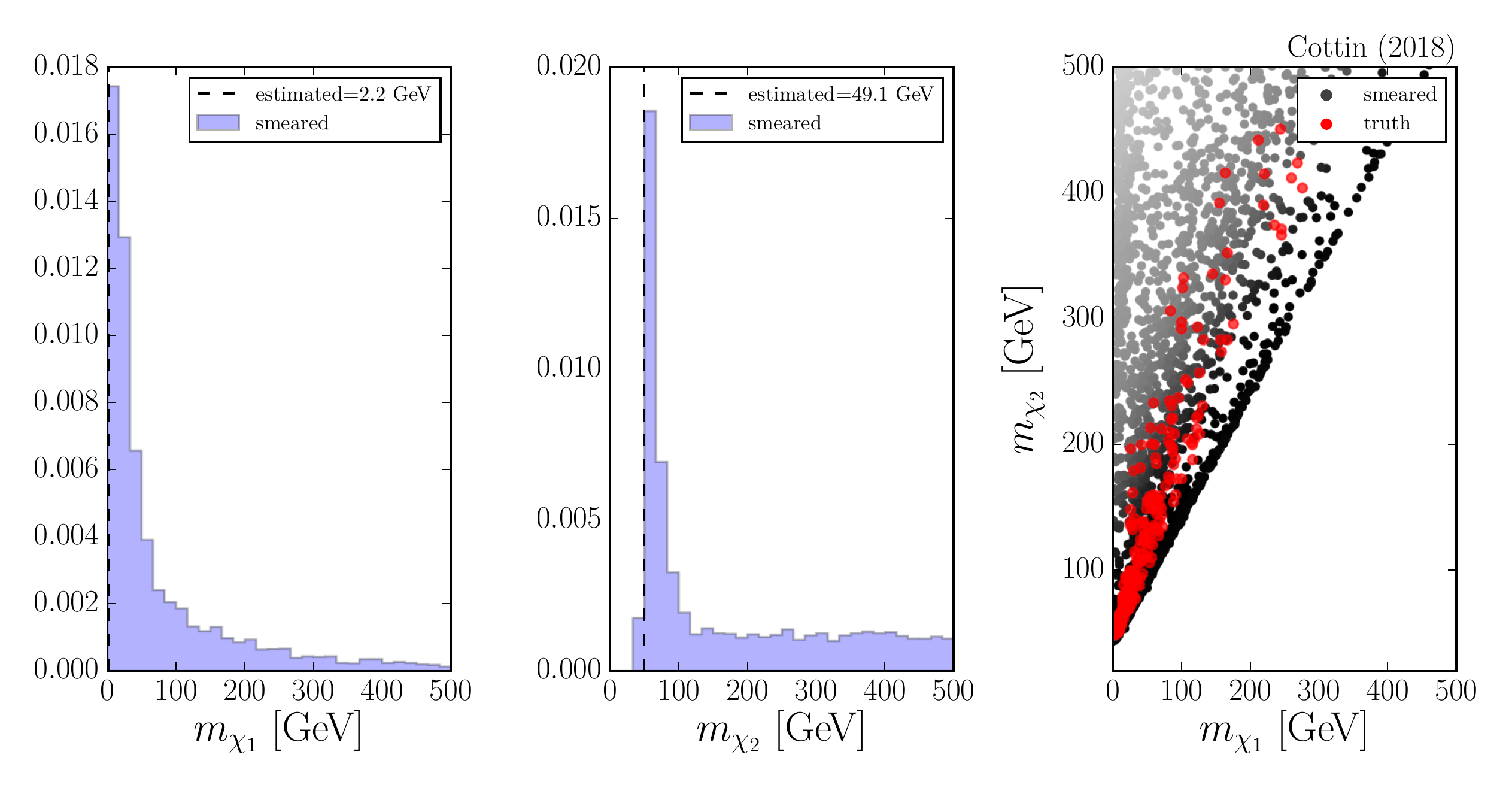}
\caption{ Solutions for the masses including all detector effects. The estimated mass values for the daughter $\chi_{1}$ (left) and parent $\chi_{2}$ (middle) are shown in the label. The right far plot shows the truth solutions and the solutions after considering all detector effects (or smearings). The truth values of the mass pair are $(m_{\chi_1}, m_{\chi_2})=(1, 50)$ GeV.}
\label{figSmeared}
\end{figure}

After smearing, we compute an {\it{estimate}} for the mass pair $(m_{\chi_1}, m_{\chi_2})$ based on a fixed percentile of the data formed with the set of solutions arising from the equation solving process. Figure~\ref{figSmeared} shows the distributions of the parent and daughter masses after smearing $\mathcal{O}(10000)$ independent events, including all detector effects. We also show the truth mass solutions in red, together with the solutions for the mass pair after considering all smearings in the far right plot. The estimated mass values are calculated from the first percentile of the data in each mass distribution. The estimated pair in this case is $(m_{\chi_1}, m_{\chi_2})= (2.2, 49.1)$ GeV for a truth mass pair of $(m_{\chi_1}, m_{\chi_2})=(1, 50)$ GeV.

We now wish to address the following question: if there are displaced events seen at colliders of the topology in Figure~\ref{MyTopology}, and since there are no SM contribution to displaced vertices of the topology assumed, how heavy are the masses of the parent and daughter particles? Basically we would like to extract both parent and daughter masses from the data. 

\begin{figure}[h]
\centering
\includegraphics[width=0.5\textwidth,angle=0]{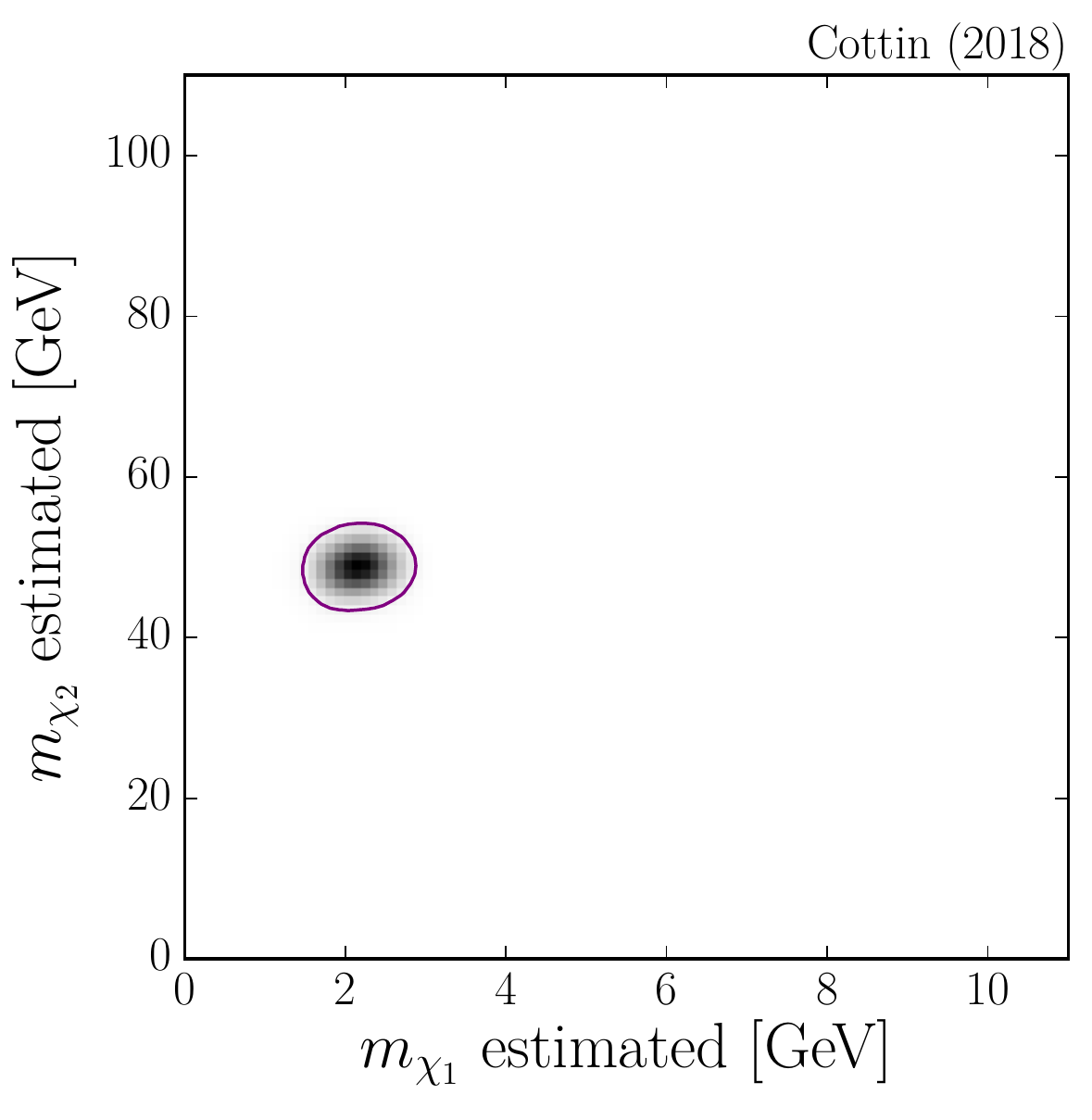}
\caption{ Two-dimensional estimator map generated for the truth values $(m_{\chi_1}, m_{\chi_2})= (1, 50)$. $95 \%$ of the points lie inside the purple contour.}
\label{figEstimated}
\end{figure}

\begin{figure}[h]
\centering
\includegraphics[width=0.5\textwidth,angle=0]{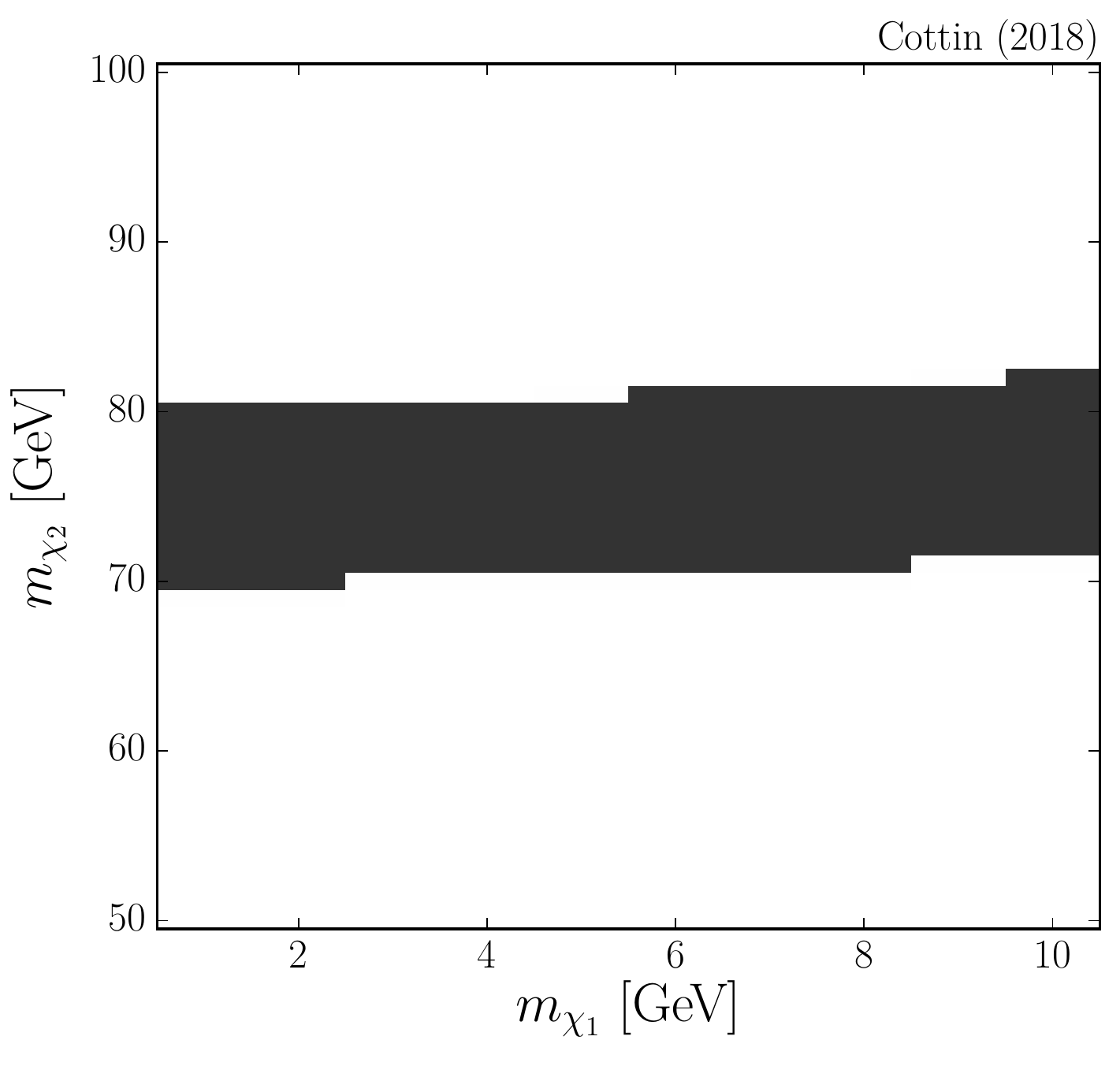}
\caption{Median $95\%$ confidence region for five observations of the mass pair $(m_{\chi_1},m_{\chi_2})$. The observations are in Table~\ref{tab:observed}. }
\label{figCR}
\end{figure}

To illustrate the sort of confidence intervals that might result from the observation of one event containing displaced vertices in this simplified model, we generate an example of a $95\%$ confidence region in $m_{\chi_1}-m_{\chi_2}$ space, using the coverage properties of a two-dimensional estimator defined, as above, by the first percentile of the distribution of solutions arising from the equation solving process. An example of the estimated masses can be seen in Figure~\ref{figEstimated}. In Figure~\ref{figCR} we show a confidence region that uses a set of five events (see Table~\ref{tab:observed}) typical of a mass regime consistent with $(m_{\chi_1}, m_{\chi_2})= (5, 75)$, thinking of the middle of the grid of points we sampled.

\begin{table}[h] 
\centering
\begin{tabular*}{
0.64\textwidth
}{ c }
\hline
      Set of observed values $(m_{\chi_{1}},m_{\chi_{2}})$    \\ 
\hline \hline 
  $
   (5.2 , 73.0),    
   (4.7 , 73.0  ),  
   (4.6 , 73.2    ),
   (5.2 , 73.4    ),
   (5.5 , 73.8  )$
   \\
\hline 
\end{tabular*} 
\caption{\label{tab:observed} A set of five observations for the mass pair $(m_{\chi_{1}},m_{\chi_{2}})$. These were randomly selected before creating the estimation maps to construct the median $95\%$ confidence region in Figure~\ref{figCR}.}
\end{table} 

\section{Conclusions}
\label{close}

We presented a simple method for reconstructing particle masses in events with displaced vertices, whose 
utility is motivated by models with displaced dark matter signatures at colliders. 

We considered event topologies where two long-lived parent particles decay to two invisible daughters and visible particles. Knowing the displaced vertex positions of the two parents, with the assumption that the momentum of the parent lies in direction of the displaced vertex, gives us extra information to constrain the kinematics fully. Note that this topology is not restricted to cases where the long-lived parent is neutral, making the method accessible to various models.

After considering detector effects in the reconstruction of displaced vertices inside the inner tracker, leptons, jets and missing transverse momentum, we produce an {\it{estimate}} for the daughter-parent mass pair. The estimate is based on the first percentile of the data formed with the set of solutions for the mass pair. After constructing mass histograms in estimation space for a mass grid in the displaced dark matter simplified model in~\cite{Buchmueller:2017uqu}, given an observation, a 95 $\%$ confidence region is constructed. 

The technique presented can be tested for other models, and can be extended to consider additional final states (such as muons or jets) to arrive at results of the kind shown in Figure~\ref{figCR}. The method can be used, provided the assumptions the technique relies on are respected and consistent detector simulation is done for these (displaced) objects. Further refinements of the method can also be made by considering additional detector effects, such a mis-reconstructions in the primary interaction point.

Looking for displaced decays will continue to be an important signature in the discovery of new physics. If displaced events are seen at the LHC, this method proves useful to constrain unknown particle masses, and may shed light on the mass scale for dark matter.

\begin{acknowledgments}
The author would like to thank Ben Gripaios and Stephen West for the initial idea of this work, Ben Allanach for comments and Nishita Desai for input on the detector simulation. G.C thanks Chris Lester for invaluable discussions, advice and the reading of this manuscript. G.C is also very grateful to Gabriel Torrealba for useful conversations and his technical support with \texttt{python}. G.C was initially funded by the postgraduate CONICYT-Chile/Cambridge Trusts Scholarship 84130011 and now is supported by the Ministry of Science and Technology of Taiwan under grant No. MOST-106-2811-M-002-035.  Support from STFC grant ST/L000385/1 is acknowledged for the use of computational resources. 
\end{acknowledgments}

\appendix

\section{Kinematic equations}

We determine the momentum components and masses of the unobserved particles by solving the system of equations that embodies the kinematic structure of Figure~\ref{MyTopology}.

\begin{align}
m^{2}_{\chi_2}&=(p_{V}+p_{\chi_1})^2\\
m^{2}_{\chi'_2}&=(p_{V'}+p_{\chi'_1})^2
\end{align}

Considering the 4 on-shell mass constrains,

\begin{align}
m^{2}_{\chi_1}&=p^{2}_{\chi_1}\\
m^{2}_{\chi_2}&=p^{2}_{\chi_2}\\
m^{2}_{\chi'_1}&=p^{2}_{\chi'_1}\\
m^{2}_{\chi'_2}&=p^{2}_{\chi'_2}
\end{align}

and assuming $m_{\chi_1}=m_{\chi'_1}$ and $m_{\chi_2}=m_{\chi'_2}$ we have,

\begin{align}
m^{2}_{\chi_1}&=p^{2}_{\chi_1}=p^{2}_{\chi'_1}\nonumber\\
m^{2}_{\chi_2}&=(p_{V}+p_{\chi_1})^2=(p_{V'}+p_{\chi'_1})^2
\label{eqPB}
\end{align}

The missing transverse momentum in the event satisfies

\begin{align}
p^{x}_{\chi_1}+p^{x}_{\chi'_1}&=p^{\text{miss}}_{x}\\
p^{y}_{\chi_1}+p^{y}_{\chi'_1}&=p^{\text{miss}}_{y}
\end{align}

Including information on the displaced vertex positions $\bm{r}$, we get extra knowledge on the direction of the mommentum of the parent $\chi_2$, as

\begin{equation}
\bm{p_{\chi_2}}=|\bm{p_{\chi_2}}|\frac{\bm{r}}{r} = |\bm{p_{\chi_2}}|\bm{\hat{r}}.
\end{equation}

From $4-$momentum conservation, we have

\begin{align}
m^2_{\chi_2} &= m^2_{\chi_1}+m^{2}_{V}+2E_{V}\sqrt{m^2_{\chi_1}+|\bm{p_{\chi_1}}|^2} -2\bm{p_{V}}\cdot\bm{p_{\chi_1}}\nonumber\\
m^2_{\chi_2} &= m^2_{\chi_1}+m^{2}_{V'}+2E_{V'}\sqrt{m^2_{\chi_1}+|\bm{p_{\chi'_1}}|^2} -2\bm{p_{V'}}\cdot\bm{p_{\chi'_1}},
\label{eqSet}
\end{align}

where the unknown quantities are $m_{\chi_2}$, $m_{\chi_1}$, $\bm{p_{\chi_1}}$ and $\bm{p_{\chi'_1}}$.\footnote{ Note that in the simpler case where $m_{\chi_1}=0$, the system of equations in~(\ref{eqSet}) together with the constrain from equation~(\ref{eqPB}) is enough to solve for $|\bm{p_{\chi_2}}|$, as shown for example in Reference~\cite{Gripaios:2012th}, where we recover the same result if we identify $m_{\chi_1}$ with the (massless) neutrino. } In order to solve the system of equations, we first define the 3-momentum of $V$ ($V'$) and $\chi_1$ ($\chi'_1$) in terms of their projections to the particle $\chi_2$ ($\chi'_2$), whose direction is known. The parallel $\parallel$ and perpendicular $\bot$ components are

\begin{align}
(\bm{p_{\chi_1}})_{\parallel \chi_2}&=(\bm{p_{\chi_1}}\cdot\bm{\hat{r}})\bm{\hat{r}}\\
(\bm{p_{V}})_{\parallel \chi_2}&=(\bm{p_{V}}\cdot\bm{\hat{r}})\bm{\hat{r}}\\
(\bm{p_{\chi_1}})_{\bot \chi_2}&=\bm{p_{\chi_1}} - (\bm{p_{\chi_1}}\cdot\bm{\hat{r}})\bm{\hat{r}}\\
(\bm{p_{V}})_{\bot \chi_2}&=\bm{p_{V}} - (\bm{p_{V}}\cdot\bm{\hat{r}})\bm{\hat{r}}.
\end{align}

Considering $(\bm{p_{\chi_1}})_{\bot \chi_2}=-(\bm{p_{V}})_{\bot \chi_2}$, we have that


\begin{equation}
\bm{p_{\chi_1}}=(A+B)\bm{\hat{r}}-\bm{p_{V}}
\end{equation}

where we have defined 

\begin{align}
A&\equiv(\bm{p_{\chi_1}}\cdot\bm{\hat{r}})\nonumber\\ 
B&\equiv(\bm{p_{V}}\cdot{\bm{\hat{r}}}). 
\label{eq:AB}
\end{align}

Similarly for $\bm{p}_{\chi'_1}$ we have

\begin{equation}
\bm{p_{\chi'_1}}=(C+D)\bm{\hat{r}'}-\bm{p_{V'}}
\end{equation}

with 

\begin{align}
C&\equiv(\bm{p_{\chi'_1}}\cdot\bm{\hat{r}'}) \nonumber\\
D&\equiv(\bm{p_{V'}}\cdot{\bm{\hat{r}'}}). 
\label{eq:CD}
\end{align}

Note that the unknown quantities are $A$ and $C$, which we can clear by using the following constrain on the missing transverse momenta $\bm{p}^{\text{miss}}_{T}=(p^{\text{miss}}_{x},p^{\text{miss}}_{y})$ in the event

\begin{equation}
\bm{p}^{\text{miss}}_{T}=[(A+B)\bm{\hat{r}}-\bm{p_{V}} + (C+D)\rhatp-\pvp]_{\bot},
\end{equation}

which allows to extract the unknown quantities,



\begin{align}
A
&=
\frac{ 
\rhatp\times(\pv+\pvp+ \bm{p}_T^{\text{miss}})\cdot \bm{k}
}
{
\rhatp\times\rhat\cdot \bm{k}
}
-\pv \cdot \hat{\bm{r}}
\label{eqA}
\\
C
&=
\frac{ 
\rhat\times(\pv+\pvp+ \bm{p}_T^{\text{miss}})\cdot \bm{k}
}
{
\rhat\times\rhatp\cdot \bm{k}
}
-\pvp\cdot \rhatp
\label{eqC}
\end{align}

where $\bm{k}$ is a fixed three-vector pointing along the beam-line.\footnote{Note that in equations (\ref{eqA}) and (\ref{eqC}) the quantity $\bm{p}_T^{\text{miss}}$ is used as if it were a three-vector. This is possible, even though $\bm{p}_T^{\text{miss}}$ has no well-defined component parallel to the beam-line, since the vector $\vec{k}$ makes (\ref{eqA}) and (\ref{eqC}) insensitive to any $z$-component it might be assigned.   } We also have that,

\begin{align}
\pchione
&=
(A+B)\rhat-\pv \nonumber
\\
&=
\left( \frac{ 
\rhatp\times(\pv+\pvp+ \bm{p}_T^{\text{miss}})\cdot \bm{k}
}
{
\rhatp\times\rhat\cdot \bm{k}
}
-\pv \cdot \rhat
+ 
\pv \cdot \rhat
\right)\rhat-\pv \nonumber
\\
&=
\left( \frac{ 
\rhatp\times(\pv+\pvp+ \bm{p}_T^{\text{miss}})\cdot \bm{k}
}
{
\rhatp\times\rhat\cdot \bm{k}
}
\right)\rhat-\pv,\qquad\text{and} \nonumber
\\
\pchionep
&=
(C+D)\bm{\hat{r}}'-\bm{p}_{V}' \nonumber
\\
&=
\left( \frac{ 
\hat{\bm{r}}\times(\bm{p}_{V}+\bm{p}_{V'}+ \bm{p}_T^{\text{miss}})\cdot \bm{k}
}
{
\hat{\bm{r}}\times\hat{\bm{r}}'\cdot \bm{k}
}
-\bm{p}'_{V} \cdot \hat{\bm{r}}'
+ 
\bm{p}'_{V} \cdot \hat{\bm{r}}'
\right)\bm{\hat{r}'}-\bm{p}'_{V} \nonumber
\\
&=
\left( \frac{ 
\hat{\bm{r}}\times(\bm{p}'_{V}+\bm{p}_{V'}+ \bm{p}_T^{\text{miss}})\cdot \bm{k}
}
{
\hat{\bm{r}}\times\hat{\bm{r}}'\cdot \bm{k}
}
\right)\bm{\hat{r}'}-\bm{p}'_{V}
\label{eq:pchiOnes}
\end{align}

or equivalently

\begin{align}
\pchitwo
&=
\pchione+\pv \nonumber
\\
&=
\left( \frac{ 
\rhatp\times(\pv+\pvp+ \bm{p}_T^{\text{miss}})\cdot \bm{k}
}
{
\rhatp\times\rhat\cdot \bm{k}
}
\right)\rhat,\qquad\text{and} \nonumber
\\
\pchitwop 
&=
\pchionep+\pvp \nonumber
\\
&=
\left( \frac{ 
\rhat\times(\pv+\pvp+ \bm{p}_T^{\text{miss}})\cdot \bm{k}
}
{
\rhat\times\rhatp\cdot \bm{k}
}
\right)\rhatp.
\label{eq:pchiTwos}
\end{align}

The last two equations are telling us the magnitudes of $\pchitwo$ and $\pchitwop$ are given by

\begin{align}
\frac{ 
\rhatp\times(\pv+\pvp+ \bm{p}_T^{\text{miss}})\cdot \bm{k}
}
{
\rhatp\times\rhat\cdot \bm{k}
}
\label{eq:max}
\end{align} and
\begin{align}
\frac{ 
\rhat\times(\pv+\pvp+ \bm{p}_T^{\text{miss}})\cdot \bm{k}
}
{
\rhat\times\rhatp\cdot \bm{k}
}
\label{eq:moritz}
\end{align} respectively.  There is no guarantee, however, that these quantities need be positive in the face of measurement uncertainties.  If either quantity were found to be negative, that would consititute evidence that either (i) the event is incompatible with the hypothesised kinematic structure, or (ii) something has been imperfectly measured.
The quantity in (\ref{eq:max}) is positive if, in the transverse plane, $\pv + \pvp + \bm{p}_T^{\text{miss}}$ is on the same side of $\rhatp$ as $\rhat$.  Likewise, the quantity in (\ref{eq:moritz}) is positive if, in the transverse plane, $\pv + \pvp + \bm{p}_T^{\text{miss}}$ is on the same side of $\rhat$ as $\rhatp$.  Together these imply that the event admits the hypothesised kinematic structure only if $\pv+\pvp+ \bm{p}_T^{\text{miss}}$ lies {\it between} $\rhat$ and $\rhatp$ in the transverse plane.  This should be no surprise; under the hypothesised kinematic structure $\pv + \pvp + \bm{p}_T^{\text{miss}}$ is the total transverse momentum of the $\chi_2$ and $\chi'_2$, and this is writable as
$|\pchitwo|\rhat+ |\pchitwop|\rhatp$ and so must lie between $\rhat$ and $\rhatp$.

Provided that 
$\pv+\pvp+ \bm{p}_T^{\text{miss}}$ lies {\it between} $\rhat$ and $\rhatp$ in the transverse plane, we have succeed in the programme of solving for $\pchitwo$, $\pchitwop$, $\pchione$ and $\pchionep$ entirely in terms of the event variables ${\bm r}$, ${\bm r'}$, $\pv$, $\pvp$ and ${\bm p}_T^{\text{miss}}$. 

\bibliographystyle{JHEP}
\bibliography{massDV}

\end{document}